# Mobility Based Routing Protocol with MAC Collision Improvement in Vehicular Ad Hoc Networks

Zhihao Ding, Pinyi Ren, Qinghe Du

Shaanxi Smart Networks and Ubiquitous Access Rearch Center
School of Electronic and Information Engineering, Xi'an Jiaotong University
dingzhihao@stu.xjtu.edu.cn

**Abstract.** Intelligent transportation system attracts a great deal of research attention because it helps enhance traffic safety, improve driving experiences, and transportation efficiency. Vehicular Ad Hoc Network (VANET) supports wireless connections among vehicles and offers information exchange, thus significantly facilitating intelligent transportation systems. Since the vehicles move fast and often change lanes unpredictably, the network topology evolves rapidly in a random fashion, which imposes diverse challenges in routing protocol design over VANET. When it comes to the 5G era, the fulfilment of ultra low end-to-end delay and ultra high reliability becomes more crucial than ever. In this paper, we propose a novel routing protocol that incorporates mobility status and MAC layer channel contention information. The proposed routing protocol determines next hop by applying mobility information and MAC contention information which differs from existing greedy perimeter stateless routing (GPSR) protocol. Simulation results of the proposed routing protocol show its performance superiority over the existing approach.

## I  Introduction

The number of vehicles has gone through an explosive growth in the last few years. Correspondingly, the increase number of vehicles has also brought serious traffic problems, such as traffic jam, gas pollution and the severe traffic safety problems. In the meanwhile, most of the modern vehicles are equipped with electronic devices like vehicular sensors, the global position system (GPS) and digital control center, which provide the possibility of building a communication network. Therefore, the emerging VANETs is aimed at providing efficient network service among vehicles[1-3]. In VANETs, vehicles can communicate with adjacent vehicles, base stations, and roadside facilities while traveling on the roads, so vehicles are able to transmit emergency messages to others which can ensure the safety of driving when there is an emergency. In most cases, vehicles are not only required to communicate with the surrounding vehicles but also with the far vehicles. Hence, the important issue in VANETs is how to build a reliable and stable multi hop communication network when vehicles are not within their communication range. There already have been many routing protocols for multi hop communications in MANET which can be divided into several categories. The existing famous topology based routing protocols are AODV, DSR, OLSR, etc [4-6]. The existing famous location based routing protocols are GPSR, GPCR, etc[7-9].

As a unique form of MANETs, VANETs has a lot of same characteristics compared to MANETs. On the one hand, they are all autonomous networks so nodes can establish wireless connections without the control of base stations. In addition, consider the constant moving status of nodes in VANETs and MANETs, so there is no have fixed network topology. Besides, VANETs differs from MANETs for the following aspects. Firstly, wireless channel in VANETs is unstable, and it is easily affected by roadside buildings, road conditions and other vehicles. Secondly, the network topology changes quickly and the link expiration time is short because of the high velocity and unstable direction of vehicles. Hence, it is a challenge to design a reliable and efficient routing protocol over VANETs.

There exists many related works on the routing protocol in VANETs. In [10], authors design a moving-zone based routing protocol by establishing moving zones that vehicles will collaborate with each other to send information in a dynamic moving zone. The authors in [11] use distance factor to minimize control overhead by selecting the least number of possible hops. Authors in [12] propose an improved GPSR routing protocol by using vehicle density and design a new data structure in order to carry route information which can be considered when search routing.

In this paper, we propose a novel routing protocol by using the information of mobility along with a MAC collision improvement. As is illustrated in [11], increased density of nodes challenges the utilization of resources, which motivates us to optimize MAC performance. We assume the network operates on the IEEE 802.11 DCF mechanism [12]. In the first step, we calculate the link expiration time using the mobility. In the second step, we estimate the delay in channel contention under 802.11 DCF mechanism and we also consider the greedy strategy. Finally, we calculate the forward weight of node and combine with GPSR routing protocol to determine the next hop.

The rest of this paper is organized as follows. We in section II introduce the proposed routing protocol with mobility and MAC collision improvement. We in section III presents the simulation results under our proposed routing protocol and GPSR routing protocol. Finally, the conclusions are derived in Section IV.

## II    The Proposed Routing Protocol

We in this section present our modified routing scheme in detail. In the beginning, our greatest wish is to reduce the transmission delay as much as possible since a lot of vehicle applications are delay sensitive. And it's also essential to enhance link connectivity due to the various speed and direction of different vehicles. We first assume the information of vehicle speed and direction can be collected by GPS and other vehicular sensors. There exists a transmission range and each vehicle can communication with each other within the transmission range. In common case, we naturally consider that routing with the shortest path is the best which will reduce the number of hops at utmost. The same idea has been used by GPSR routing protocol which can be called the greedy forwarding strategy. However, the actual situation is not always the case and we take Fig.1 for example. The source vehicle 1 wants to transmit packets to destination vehicle 10. The path obtained by GPSR routing protocol is [v1, v5, v6, v10] and it needs three hops. Recall the DCF mechanism in 802.11p, vehicles in the same cluster must experience a MAC layer contention process to use the wireless channel and every time just one packet can be sent. We can see that vehicle 6 has five neighbors in its communication range so it might experience a serious MAC delay than vehicle 8 which only has two neighbors. Therefore, this path may spend more time than the path [v1, v5, v8, v9, v10] which will spend four hops to arrive the destination. Now we first estimate the MAC delay by analyzing the contention process of the MAC layer.

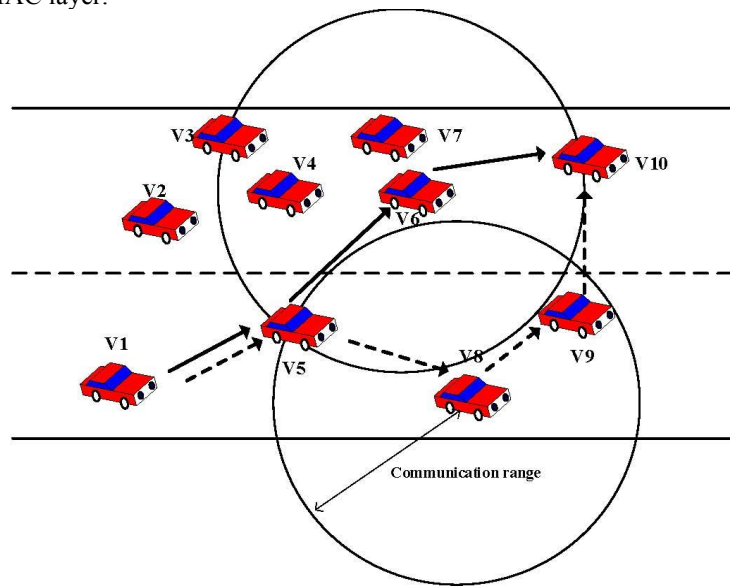

**Figure 1 An example of different routing selections**

### A. MAC Delay Analysis

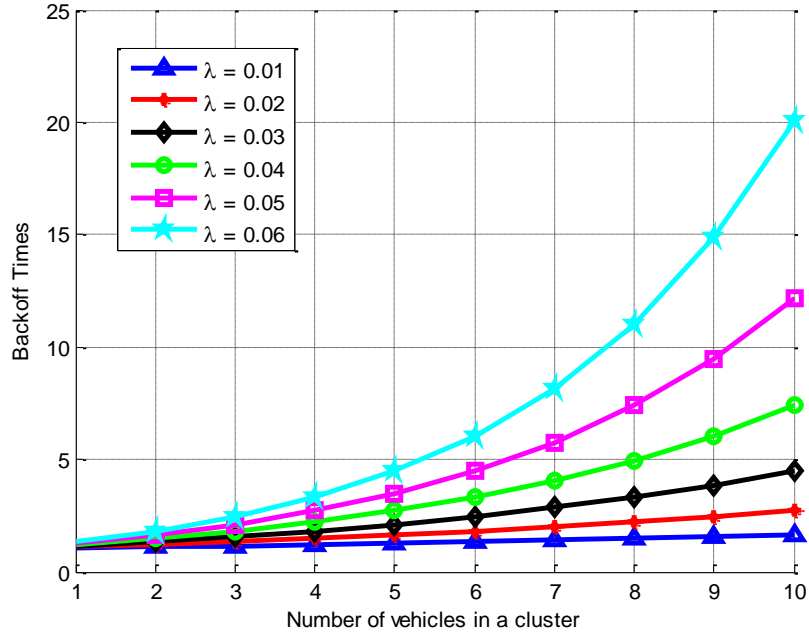

**Figure 2 Average backoff times in a cluster under different packet arrival rate**

IEEE 802.11p is the standard protocol in VANETs which illustrates the main technical standards for the PHY layer and MAC layer [13]. The basic scheme is DCF which known as carrier sense multiple access with collision avoidance (CSMA/CA). In this situation, vehicles must sense the channel state before transmitting packets. The contention process in MAC layer can be simplified as follows. Firstly, vehicle initiates a random backoff procedure until the medium is idle for DIFS (DCF Inter-Frame Space) time. If the medium is busy, it must wait and continue the backoff procedure until the medium is idle once again. When the backoff time equals a zero value, a vehicle should first send a RTS (Request to Send) packet, then vehicle wait a CTS (clear to Send) packet in order to make sure the contention is success. Once a vehicle receives the CTS packet in a slot time, the contention is success and it can transmit a packet immediately. On the contrary, if no CTS packet received or vehicle detects a collision within a slot time, the vehicle turns back to the backoff state and start a new round of backoff. Usually, the common algorithm of backoff procedure in CSMA/CA is well-known BEB (Binary Exponential Backoff) algorithm [14]. In BEB, the backoff window doubles when the contention failed and there exists a minimum contention window and a maximum contention window. The backoff time is generated randomly from the window. Through the introduction to the contention process above, we can see that the number of backoff times need to be reduced as little as possible in order to reduce the delay.

In the backoff state, we assume $P_i(t)$ represents the probability that vehicle $i$ senses channel idle for time interval $t$. Therefore, $P_i(slot)$ can represent the success contention probability of vehicle $i$ after the backoff state when sensing channel idle for a slot time. Correspondingly, the failure probability is $1 - P_i(slot)$. Hence, the probability of $k$-th backoff before a successful transmission is

$$N(K) = (1 - P_i(slot))^{k-1} \times P_i(slot) \tag{1}$$

Then, we can get the average backoff times of vehicle $i$ as follows:

$$\overline{N_i} = \lim_{n \to +\infty} \sum_{k=1}^{n} k \times (1 - P_i(slot))^{k-1} \times P_i(slot) \tag{2}$$

Now, we will calculate the probability $P_i(slot)$. The arrival of a packet of a vehicle can be seen as a Poisson process and the packet arrival process can be seen follow the Poisson distribution [15]. The probability that $n$ packets arrive at vehicle $i$ within a time interval $t$ can be expressed as

$$P_i(t, n) = \frac{(\lambda t)^n}{n!} e^{-\lambda t} \tag{3}$$

where $\lambda$ denotes the packet arrive rate of a vehicle.

We notice that the overall packet arrival rate $\lambda$ in cluster $j$ is

$$\lambda(j) = C(j) \times \lambda \tag{4}$$

where $C(j)$ denotes the whole number of vehicles in cluster $j$.

Consequently, for vehicle $i$ in cluster $j$, the probability that no vehicle else to transmit packets during time interval $t$ is written as follows:

$$P_i(t, 0) = e^{-\lambda(j) \times t} \tag{5}$$

Then we can solve this limit expression. For the sake of simplicity, we first ignore the limit symbol and replace formula (5) to formula (2). The expression can be transformed as follows:

$$\overline{N_i} = \lim_{n \to +\infty} \sum_{k=1}^{n} k \times (1 - e^{-\lambda(j) \times t})^{k-1} \times e^{-\lambda(j) \times t} \tag{6}$$

If the left side of formula (6) multiplied by $1 - e^{-\lambda(j) \times t}$, we can get a new formula as follows:

$$(1 - e^{-\lambda(j) \times t}) \times \overline{N_i} = \lim_{n \to +\infty} \sum_{k=1}^{n} k \times (1 - e^{-\lambda(j) \times t})^{k} \times e^{-\lambda(j) \times t} \tag{7}$$

Formula (6) minus Formula (7) and results can be simplified as follows:

$$\overline{N_i} = \sum_{k=1}^{n} k \times (1 - e^{-\lambda(j) \times t})^{k-1} - n(1 - e^{-\lambda(j) \times t})^n \tag{8}$$

The first term of formula (8) is geometric series and the second term of formula (8) can be transformed into the problem of solving the limit of continuous function. Obviously, the second term equals zero. Finally, when $n$ tends to positive infinity and we can get the average backoff times of vehicle $i$ in cluster $j$ as follows:

$$\overline{N_i} = \frac{1}{e^{-C(j) \times \lambda \times slot}} \tag{9}$$

Fig.2 shows the average backoff times of a vehicle under different $\lambda$ and different number of vehicles in a cluster. It's obvious that the less the number of backoff, the lower the MAC layer delay. So we should try to choose the vehicle with fewer neighbor nodes as next hop.

**B. Link Expiration Time**

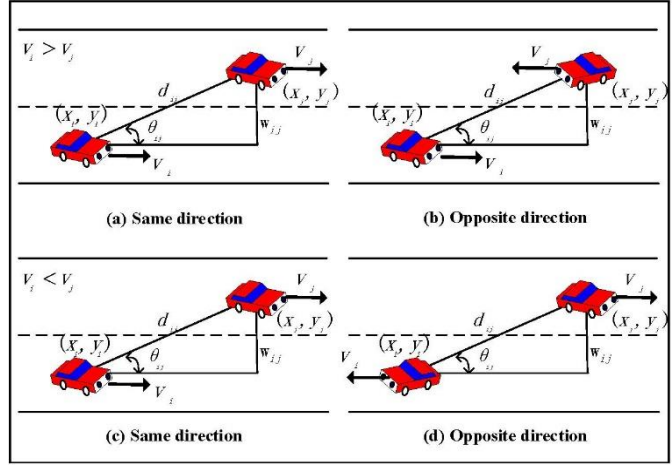

Figure 3 An example of calculating LET under different vehicle speed and direction; (a)Same direction with $v_i > v_j$. (b)Opposite direction. (c)Same direction with $v_i < v_j$. (d)Opposite direction.

We assume that the mobility information including vehicle speed and direction can be obtained by GPS device. We also assume each vehicle can collect mobility information of others which within the communication range. Then we can estimate the lifetime of link by calculating LET. LET measures the reliability of a link. The communication range of each vehicle is set to $R$. According to the moving situations of vehicles on the road, the LET can be expressed into four different conditions which shown in Fig.3. We assume the speed of vehicle $i$ and vehicle $j$ are $v_i$, $v_j$ respectively. The coordinates of the two vehicles are $(x_i, y_i)$ and $(x_j, y_j)$. Therefore, the distance between vehicle $i$ and vehicle $j$ is

$$d_{ij} = \sqrt{(x_i - x_j)^2 + (y_i - y_j)^2} \qquad (10)$$

The angle $\theta_{ij}$ is

$$\theta_{ij} = \arctan \frac{y_i - y_j}{x_i - x_j} \qquad (11)$$

Then we can get the vertical distance of vehicle $i$ and vehicle $j$:

$$w_{ij} = d_{ij} \sin \theta_{ij} \qquad (12)$$

Consequently, when vehicle $i$ and vehicle $j$ are moving in the same direction, the LET between two vehicles is as follows:

$$LET_{ij} = \frac{\sqrt{R^2 - w_{ij}^2} \pm \sqrt{d_{ij}^2 - w_{ij}^2}}{|v_i - v_j|} \quad (13)$$

when vehicle $i$ and vehicle $j$ are moving in the opposite direction, the LET between two vehicles is as follows:

$$LET_{ij} = \frac{\sqrt{R^2 - w_{ij}^2} \pm \sqrt{d_{ij}^2 - w_{ij}^2}}{|v_i + v_j|} \quad (14)$$

Now, we have estimated the MAC backoff times and LET of each vehicle, then we modify the next hop selection strategy of GPSR routing protocol. We assume each vehicle has a node weight and for a source vehicle, the weight of neighbor $i$ is calculated as follows:

$$W_i = \alpha \frac{\overline{N_{max}} - \overline{N_i}}{\overline{N_{max}}} + \beta \frac{LET_{ij}}{LET_{max}} + \gamma \frac{d_{max} - d_i}{d_{max}} \quad (15)$$

where $LET_{max}$ is the maximum LET of all neighbors; $LET_{ij}$ represents the LET between vehicle $i$ and source vehicle. $\overline{N_{max}}$ is the maximum backoff times of all neighbors; $\overline{N_i}$ is the backoff times of vehicle $i$; $d_{max}$ denotes the maximum distance between neighbors and destination; $d_i$ is the distance between vehicle $i$ and destination; $\alpha + \beta + \gamma = 1$.

Obviously, we choose the vehicle as next hop which has the maximum value of $W_i$. The procedure of the proposed routing protocol can be summarized for following process. Firstly, the source vehicle measure the distance between itself and destination vehicle. When these two vehicles are within their communication range $R$, source vehicle can tramsmit packet to the destination vehicle directly. If the distance is more than $R$, we calculate the LET, backoff times and distance of each neighbor vehicle. Then we can get the value $W_i$ of each neighbor vehicle and choose the vehicle with maximum value of $W_i$ as forward vehicle. And we repeat this process to find the whole path. The detailed procedure of route discovery is presented in Algorithm 1.

A route maintenance strategy is considered in our proposed routing protocol as well in order to avoid link disconnection. The link disconnection is more common in VANETs because of the influence of vehicle speeds and directions. The idea of route maintenance is to periodically detect the distance between vehicles in order to determine whether or not it is interrupted. The detailed procedure of route maintenance is presented in Algorithm 2.

## Algorithm 1 Route Discovery

1. Get speed and direction information of all vehicles;
2. Calculate the distance between current transmit vehicle and all other vehicles;
3. Get the neighbor vehicles to neighbor set M;

4. n =0;
5. **while** n < M.size **do**
6.    **if** vehicle n is destination vehicle
7.      Select destination vehicle;
8.      Return;
9.    **else**
10.      Apply Eq.(5) to calculate $W_i$
11.    **end if**
12.    n = n + 1;
13. **endwhile**
14.    **Select the vehicle with maximum value of weight** $W_i$

**Algorithm 2 Route Maintance**

1 Update position of all vehicles;
2 Get all source vehicle in set S;
3 n =0;
4   **while** current vehicle isn't destination vehicle and n < S.size **do**
5     Calculate the distance d between current vehicle and next hop vehicle;
6    **if** d < 250
7     Let current vehicle be next hop vehicle
8    **else**
9     Apply current vehicle to route discovery procedure;
10     c
11    **end if**
12    **if** current vehicle is destination vehicle
13     n = n + 1;
14    **end if**
15   **endwhile**

## III    Performance Evaluations

In this section, our proposed routing protocol is evaluated and compared with the famous GPSR routing protocol. We consider a highway model environment with vehicles are distributed randomly for simplicity. The highway model environment is 1000m length and has six lanes in two directions. The wireless communication range is set to 250m. The velocity of vehicles is variable from 30km/h to 80km/h. Table 1 presents the detailed parameters in our simulation.

| Parameters | Value |
|---|---|
| Simualtion scenario | Highway |
| Number of lanes | 6 |
| Simalation area | 1000m*25m |
| Data packet size | 512bytes |
| Cwminimun | 31 |
| CWmaximun | 1023 |
| $\alpha$ | 1/3 |
| $\beta$ | 1/3 |
| $\gamma$ | 1/3 |
| Arrival rate ($\lambda$) | 10packets/ms |
| Speed of vehicles | 30-80km/h |
| Number of vehicles | 12-60 |
| Communication range | 250m |
| MAC protocol | IEEE802.11p |
| Routing protocol | The proposed, GPSR |
| Simulation time | 500TTI |

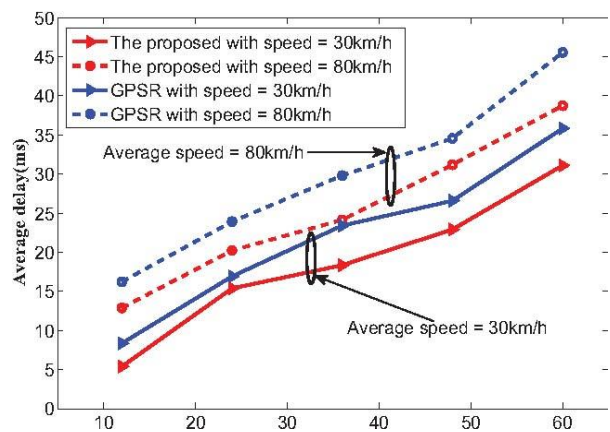

**Figure 4 Average end-to-end delay vs number of vehicles**

Fig.4 illustrates the average end-to-end delay of our proposed routing protocol and GPSR routing protocol versus the number of vehicles in different average vehicle speed. End-to-end delay refers to the time that packets transmitted among two vehicles plus the delay in MAC layer. From the figure, we firstly see that the delay increases as the vehicle density increases both in GPSR routing protocol and our proposed routing protocol. It's reasonable because the contention of MAC layer will become more competitive while the number of vehicles increases and the time cost in contention becomes bigger. Secondly, when the speed of vehicle increases, the delay will increase accordingly as high speed leads to link instability. Thirdly, end-to-end delay under our proposed routing protocol is smaller than GPSR routing protocol. It's expected for the following reason. As the mobility information and MAC delay estimation are considered in our proposed routing protocol, the choice of next hop will become more reasonable than GPSR which only considers the distance. Thus, our proposed routing protocol performs better than GPSR on end-to-end delay.

Fig.5 plots the packet delivery rate of our proposed routing protocol and GPSR routing protocol versus the number of vehicles in different average vehicle speed. From this figure, we can see packet delivery rate of both two routing protocols increases when the density of vehicles increases. This is obviously because we can choose a more appropriate vehicle node for packet forwarding when there are more vehicles. Besides, when the number of vehicles is less, there are few effective paths can be selected so the packet delivery rate is similar to that of the two protocols. We can also see that packet delivery rate gets worse with the continuous increase of vehicle speed which is expected. When the number of vehicles is large, the path in our proposed routing protocol is more reliable, so the packet delivery rate is higher than GPSR routing protocol.

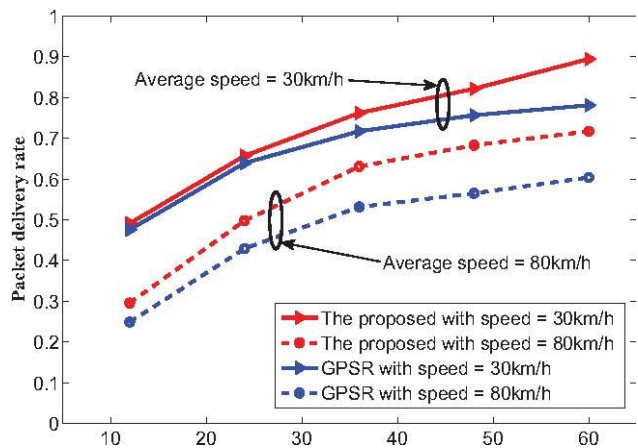

**Figure 5 Packet delivery rate vs number of vehicles**

Fig.6 illustrates the broken links against the number of vehicles in different average vehicle speed. The metrics of broken links measures the link reliability. If the number of broken links is less, the route can be considered stable and reliable. Otherwise, if wireless links interrupt frequently, the overheads of exchanging control packets will become larger and it will spend some delay in reestablishing links.

Additionally, this figure demonstrates that the number of broken links decreases as the number of vehicles increases and the link interruption is more serious when the speed is 80km/h. Finally, the number of broken links of our proposed routing protocol is small compared to GPSR routing protocol which illustrates the proposed routing protocol is more stable once again.

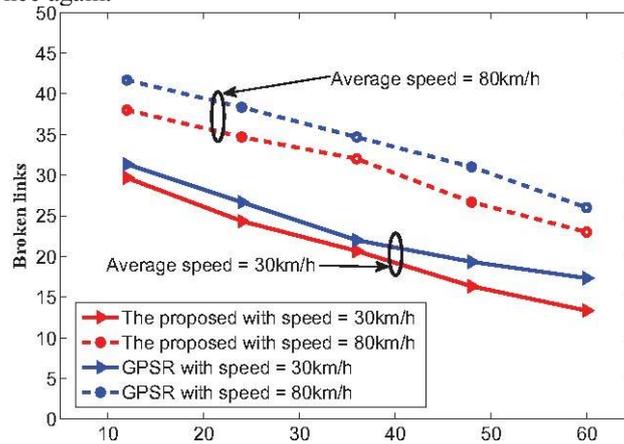

**Figure 6 Broken links vs number of vehicles**

## IV    Conclusions

This paper described an improved routing protocol in VANETs by modifying the forward procedure of existing GPSR routing protocol. The highlight of our proposed routing protocol is that we reduce the delay and strengthen the reliablity of routes. We optimize the next hop selection strategy of the well-known GPSR routing protocol by using mobility and MAC delay estimation. The method in our routing protocol is based on the vehicular electronic devices which can provide a large amount of vehicular information. We add mobility information(speed and direction) and MAC delay estimation in next hop selection strategy to make the route more reliable and decrease delay. We also discuss the route maintenance strategy in detail. Finally, Abundant and convincing simulations prove that our proposed routing protocol performs much better in forms of average end-to-end delay, packet delivery rate and broken links than GPSR.

## References

1. Y. H. Zhu, X. Z. Tian and J. Zheng, ``Routing in vehicular ad hoc networks: A survey,'' *IEEE Vehicular Technology Magazine*, vol. 2, no. 2, pp. 12--22, Jun. 2007.
2. M. Altayeb, I. Mahgoub, ``A survey of vehicular ad hoc networks routing protocols, *International Journal on Innovation and Applied Studies* , vol. 3, no. 3, pp. 829--846, Jun. 2013.

3. M. Azees, P. Vijayakumar and L. Jegatha Deborah, ``Comprehensive survey on security services in vehicular ad-hoc networks,'' *IET Intelligent Transport Systems*, vol. 10, no. 6, pp. 379--388, Aug. 2016.
4. J. J. Ferronato and M. A. S. Trentin, ``Analysis of routing protocols OLSR, AODV and ZRP in real urban vehicular scenario with density variation,'' *IEEE Latin America Transactions*, vol.15, no. 9, pp.1727--1734, Aug. 2017.
5. R. Bai and M. Singhal, ``DOA: DSR over AODV routing for mobile ad hoc networks,'' *IEEE Transactions on Mobile Computing*, vol. 5, no. 10. pp. 1403--1416, Oct. 2006.
6. J. Toutouh, J. Garcia-Nieto and E. Alba, ``Intelligent OLSR routing protocol optimization for VANETs,'' *IEEE Transactions on Vehicular Technology*, vol. 61, no. 4, pp. 1884--1894, May 2012.
7. C. Tripp Barba, L. Urquiza Aguiar and M. Aguilar Igartua, ``Design and evaluation of GBSR-B, an improvement of GPSR for VANETs,'' *IEEE Latin America Transactions*, vol.11, no. 4, pp. 1083--1089, Jun. 2013.
8. C. Lochert, M. Mauve and H. Hartenstein, ``Geographic routing in city scenarios,'' *ACM Sigmobile Mobile Computing and Communications Review,* vol. 9, no. 1, pp. 69--72, Sep. 2005.
9. B. Karp and H. T. Kung, ``GPSR: greedy perimeter stateless routing for wireless networks,'' International Conference on Mobile Computing and Networking, pp. 243--254, 2000.
10. D. Lin, J. Kang, A. Squicciarini, et.al. ``MoZo: A moving zone based routing protocol using pure V2V communication in VANETs,'' *IEEE Transactions on Mobile Computing*, vol. 16, no. 5, pp. 1357--1370, May 2017.
11. Y. R. Bahar, N. F. Abdullah, M. Ismail, et.al. ``Efficient routing algorithm for VANETs based on distance factor,'' 2016 International Conference on Advances in Electrical, Electronic and Systems Engineering (ICAEES), pp. 567--571, Mar. 2016.
12. D. Xiao and L. Asogwa, ``An improved GPSR routing protocol,'' International Journal of Advancements in Computing Techonology, vol. 3, no. 5, pp. 132--139, Jul. 2011.
13. X. Tang, P. Ren, F. Gao and Q. Du, ``Interference-Aware Resource Competition Toward Power-Efficient Ultra-Dense Networks,'' *IEEE Transactions on Communications*, vol. 65, no. 12, pp. 5415-5428, Dec. 2017.
14. T. Sakurai and H. L. Vu, ``MAC access delay of IEEE 802.11 DCF,'' *IEEE Transactions on Wireless Communications*, vol. 6, no. 5, pp. 1702--1710, May 2007.
15. C. Y. Chang, H. C. Yen and D. J. Deng, ``V2V QoS guaranteed channel access in IEEE 802.11p VANETs,'' *IEEE Transactions on Dependable and Secure Computing*, vol. 13, no. 1, pp. 5--17, Jan. 2016.
16. Y. H. Zhu, X. Z. Tian and J. Zheng, ``Performance Analysis of the Binary Exponential Backoff Algorithm for IEEE 802.11 Based Mobile Ad Hoc Networks,'' 2011 IEEE International Conference on Communications (ICC), pp. 1--6, May 2011.
17. S. Sheu and J. Chen, ``A novel delay-oriented shortest path routing protocol for mobile ad hoc networks,'' IEEE International Conference on Communications, pp. 1930--1934, Aug. 2001.